\documentclass[twocolumn,english,prl,superscriptaddress,showpacs,longbibliography,fixfloat]{revtex4-1}
\usepackage[latin9]{inputenc}
\usepackage{amsmath}
\usepackage{amssymb}
\usepackage{graphicx}

\makeatletter

\pdfpageheight\paperheight
\pdfpagewidth\paperwidth

\usepackage{xcolor}
\definecolor{darkblue}{rgb}{0.1,0.2,0.6} 
\definecolor{lightblue}{rgb}{0.1,0.1,1.0}
\definecolor{darkred}{rgb}{0.8,0.1,0.2}
\usepackage[colorlinks,citecolor=lightblue,linkcolor=darkblue,urlcolor=lightblue] {hyperref}
\renewcommand{\BibitemShut}[1]{}

\makeatother

\usepackage{babel}
\begin{document}
\global\long\def\E{\mathrm{e}}
\global\long\def\D{\mathrm{d}}
\global\long\def\I{\mathrm{i}}
\global\long\def\mat#1{\mathsf{#1}}
\global\long\def\vec#1{\mathsf{#1}}
\global\long\def\cf{\textit{cf.}}
\global\long\def\ie{\textit{i.e.}}
\global\long\def\eg{\textit{e.g.}}
\global\long\def\vs{\textit{vs.}}
 \global\long\def\ket#1{\left|#1\right\rangle }

\global\long\def\etal{\textit{et al.}}
\global\long\def\tr{\text{Tr}\,}
 \global\long\def\im{\text{Im}\,}
 \global\long\def\re{\text{Re}\,}
 \global\long\def\bra#1{\left\langle #1\right|}
 \global\long\def\braket#1#2{\left.\left\langle #1\right|#2\right\rangle }
 \global\long\def\obracket#1#2#3{\left\langle #1\right|#2\left|#3\right\rangle }
 \global\long\def\proj#1#2{\left.\left.\left|#1\right\rangle \right\langle #2\right|}

\title{Anomalous thermalization and transport in disordered interacting
Floquet systems}

\author{Sthitadhi Roy}
\email{sthitadhi.roy@chem.ox.ac.uk}

\affiliation{Physical and Theoretical Chemistry, Oxford University, South Parks
Road, Oxford, OX1 3QZ, United Kingdom}

\affiliation{Rudolf Peierls Centre For Theoretical Physics, Oxford University,
1 Keble Road, Oxford OX1 3NP, United Kingdom}

\affiliation{Max-Planck-Institut für Physik komplexer Systeme, 01187 Dresden,
Germany}

\author{Yevgeny Bar Lev}
\email{yevgeny.barlev@weizmann.ac.il}

\affiliation{Department of Condensed Matter Physics, Weizmann Institute of Science,
Rehovot 76100, Israel}

\affiliation{Max-Planck-Institut für Physik komplexer Systeme, 01187 Dresden,
Germany}

\author{David J. Luitz}
\email{david.luitz@tum.de}

\affiliation{Department of Physics, T42, Technische Universität München, James-Franck-Straße
1, D-85748 Garching, Germany}
\begin{abstract}
Local observables in generic periodically driven closed quantum systems
are known to relax to values described by periodic infinite temperature
ensembles. At the same time, ergodic static systems exhibit anomalous
thermalization of local observables and satisfy a modified version
of the eigenstate thermalization hypothesis (ETH), when disorder is
present. This raises the question, how does the introduction of disorder
affect relaxation in periodically driven systems? In this work, we
analyze this problem by numerically studying transport and thermalization
in an archetypal example. We find that thermalization is anomalous
and is accompanied by subdiffusive transport with a disorder dependent
dynamical exponent. Distributions of matrix elements of local operators
in the eigenbases of a family of effective time-independent Hamiltonians,
which describe the stroboscopic dynamics of such systems, show anomalous
departures from predictions of ETH signaling that only a modified
version of ETH is satisfied. The dynamical exponent is shown to be
related to the scaling of the variance of these distributions with
system size.
\end{abstract}
\maketitle
\emph{Introduction}.\textemdash Recent advances in cold-atom \cite{Bloch2012,ChristianGross2017}
and trapped-ion \cite{Blatt2012} experiments have stimulated interest
in nonequlibrium dynamics and thermalization or lack thereof in \emph{isolated}
quantum systems. Thermalization in both classical and quantum systems
requires an effective loss of information contained in the initial
state of the system. For classical systems, this occurs naturally,
since the underlying equations of motion are nonlinear and therefore
typically chaotic. For quantum systems, the situation is more delicate,
since all the information about the state of the system is encoded
in the wavefunction which evolves under the \emph{linear} Schrödinger
equation. While the information about the whole system cannot be lost
under unitary evolution, this is not the case for subsystems as the
\emph{corresponding reduced} density matrices evolve non-unitarily.
Thus, the objects of interest in the context of thermalization are
local observables supported on a subsystem while the rest of the system
serves as an effective bath. Written in the eigenbasis of the Hamiltonian,
the diagonal elements of local observables encode information about
the stationary state, while the offdiagonal elements contain dynamical
information. The probability distributions of diagonal and offdiagonal
matrix elements of local operators were studied already in the 80s
\cite{Peres1984,Peres1984a,Feingold1984,Feingold1985,Feingold1986},
but regained interest after the introduction of the ``eigenstate
thermalization hypothesis'' (ETH) by Deutsch \cite{Deutsch1991}
and Srednicki \cite{Srednicki1995} in the following decade. ETH was
confirmed to hold in a variety of systems \cite{Rigol2008,DAlessio2015,Borgonovi2016,Mondaini2016},
and is concerned with only the first and the second moments of the
distribution of the matrix elements, implicitly assuming that the
probability distribution is Gaussian. This assumption was motivated
by Berry's conjecture, which states that eigenstates of nonintegrable
quantum systems are reminiscent of random states drawn from a Gaussian
distribution \cite{Berry1977,Srednicki1994}. More specifically, ETH
requires that the matrix elements of local operators are described
by a smooth functional dependence on the extensive energy and the
distance from the diagonal, superimposed by random Gaussian fluctuations.
The variance of the fluctuations is assumed to decay exponentially
with the system size \cite{Srednicki1995}, which was verified for
a number of generic quantum systems \cite{Steinigeweg2013,Alba2015,Ikeda2013,Beugeling2014,Beugeling2015,Luitz2016,Mondaini2017a}.

In a recent Letter, it was shown that for a class of disordered \emph{ergodic}
systems, which for sufficiently strong disorder undergo the many-body
localization (MBL) transition \cite{Basko2006a} (see Refs.~\cite{Nandkishore2014,Abanin2017,Alet2017}
for recent reviews), ETH has to be modified, since the decay of the
fluctuations of the offdiagonal matrix elements acquires a power law
correction to their scaling with system size \cite{Luitz2016b,Serbyn2016a}.
This is accompanied with anomalous (subdiffusive) relaxation to equilibrium,
a situation which was dubbed \emph{anomalous thermalization} \cite{Luitz2016b}. 

In this work, we show that a similar phenomenology exists also in
disordered periodically driven (Floquet) systems, which undergo the
Floquet-MBL transition for sufficiently strong disorder \cite{Lazarides2014,Ponte2014,Abanin2014,Bairey2017}.
The stroboscopic dynamics of these systems is goverened by the unitrary
Floquet operator, which is the time-evolution operator over one period.
The Floquet operator can be expressed in terms of a family of effective
Hamiltonians, which allows the generalization of the concept of thermalization
and ETH to this time-dependent case \cite{DAlessio2013,Lazarides2014a,Lazarides2014b,DAlessio2014}.
It was shown that ETH assumes a simplified form since the smooth part
of the diagonal matrix elements is constant and corresponds to the
trace of the local observable \cite{DAlessio2013,Lazarides2014a,Lazarides2014b,DAlessio2014}.
This is consistent with the expectation that for any generic initial
state, the system heats up to a state which is locally indistinguishable
from the infinite temperature state. In previous studies it was numerically
shown that disordered Floquet systems exhibit anomalous heat absorption
from the periodic drive, \cite{Gopalakrishnan2015a,Gopalakrishnan2016,Kozarzewski2016a,Rehn2016}
and it was suggested that spin transport in such systems is anomalous
\cite{Gopalakrishnan2015a,Gopalakrishnan2016}. In this work using
numerically exact methods we study the nature of transport and thermalization
in disordered Floquet systems. 

\textit{Model.\textemdash }We numerically investigate a disordered
one-dimensional Heisenberg model subject to a periodic modulation
of the staggered magnetization. The driving protocol we consider is
generated by two alternating non-commuting Hamiltonians, $\hat{H}_{\pm}$
(each applied for half a period, $T/2$),

\begin{equation}
\hat{H}_{\pm}=\sum_{i=1}^{L-1}J\hat{\boldsymbol{S}}_{i}\cdot\hat{\boldsymbol{S}}_{i+1}+\sum_{i=1}^{L}\left[h_{i}\pm\left(-1\right)^{i}\Delta\right]\hat{S}_{i}^{z},\label{eq:hamfloquet}
\end{equation}
where $L$ is the length of the chain, $\hat{S}_{i}^{x,y,z}$ are
spin-1/2 operators, $J$ is the spin-spin interaction (which we set
to unity), $h_{i}\in\left[-W,W\right]$ are independent random fields
drawn from a uniform distribution, and $\Delta$ is the driving amplitude.
Both Hamiltonians commute with the total magnetization, $\hat{M}_{z}=\sum_{\ell}\hat{S}_{\ell}^{z}$,
and in what follows, we work in the $M_{z}=0$ sector for even $L$
and $M_{z}=1/2$ for odd $L$. We set $T=3$, such that the frequency
of the drive is well below the many-body bandwidth (of the undriven
system) for all considered system sizes, and choose the driving amplitude
to be small enough to still have a Floquet-MBL transition for sufficiently
strong disorder (eg. Refs.~\cite{Lazarides2014,Ponte2014}), yet
not much smaller than the singe-particle bandwidth ($\Delta=0.5$). 

\begin{figure}
\includegraphics[width=1\columnwidth]{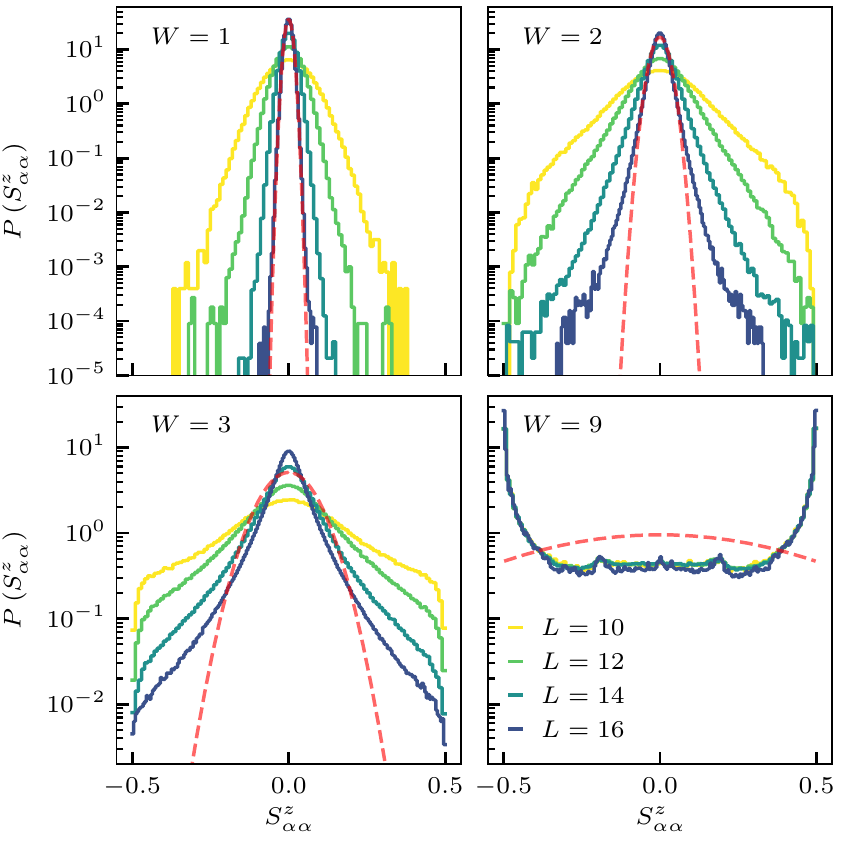}
\caption{Probability distribution, $P\left(S_{\alpha\alpha}^{z}\right)$ for
$W=1$, $2$, $3,$ $9$ and $L=10$, $12$, $14$ and $16$. The red dashed line in each panel depicts a Gaussian distribution with the same standard deviation as
that of the corresponding $P(S^z_{\alpha,\alpha})$ for $L=16$. Statistics
is obtained combining all Floquet eigenstates and all disorder realizations
(over 500 realizations for $L<16$ and 50 realizations for $L=16$).
Other parameters are $J=1$, $\Delta=0.5$ and $T=3$.}
\label{fig:fig1-diag_matel_dist} 
\end{figure}
\textit{Distributions of matrix elements.}\textemdash In Floquet systems,
the Hamiltonian is time-dependent, therefore for the study of thermalization,
the quantity of interest is not the Hamiltonian but the unitary Floquet
operator, $\hat{U}_{F}\left(T,0\right)$, which we take to be, $\hat{U}_{F}\left(T,0\right)=\E^{-\I\hat{H}_{+}T/4}\E^{-\I\hat{H}_{-}T/2}\E^{-\I\hat{H}_{+}T/4}$
\footnote{This operator is symmetric under time-reversal and has real eigenvectors,
which gives a technical advantage, since only a real eigenproblem
has to be solved.}. Using full diagonalization, we obtained all the eigenstates (denoted
by Greek letters) of $\hat{U}_{F}\left(T,0\right)$ for various system
sizes and computed the matrix elements of the local magnetization,
$S_{\alpha\beta}^{z}\equiv\left\langle \alpha\left|\hat{S}_{L/2}^{z}\right|\beta\right\rangle $.
For clean Floquet systems, it was shown that the smooth part of the
diagonal matrix elements of local operators does not depend on the
quasienergy, with fluctuations which decrease with the system size
\cite{Lazarides2014b}. We verified that this also holds for the \emph{disordered}
system we consider here \cite{SuppMat2018}. We note in passing, that
our finding rules out the existence of a mobility edge in the quasienergy
spectrum, since assuming its existence would imply that for the localized
states $\ket{\alpha}$, $\left(S_{\alpha\alpha}^{z}\right)^{2}\approx1/4$,
while for the delocalized $\left(S_{\alpha\alpha}^{z}\right)^{2}\approx0$,
which is not consistent with our numerical observation \cite{SuppMat2018}. 

While ETH is concerned only with the first and second moments of the
distributions of matrix elements, we study the full distribution of
both diagonal and offdiagonal matrix elements. The independence of
the diagonal elements on the quasienergy allows us to accumulate statistics
not only over different disorder realizations but also across all
quasienergies. To eliminate correlations, all statistical errors are
computed by bootstrap resampling over the disorder realizations only. 

In Fig.~\ref{fig:fig1-diag_matel_dist} we show the distributions
computed for various disorder strengths $W$ and system sizes $L$.
For weak disorder ($W\approx1$) the distributions are very close
to Gaussian with variances which decrease with $L$, indicating the
validity of ETH. For sufficiently strong disorder, $W=9$, the distributions
become bimodal, and almost independent of the system size, signaling
the failure of ETH which occurs in the Floquet-MBL phase. The most
fascinating situation occurs for intermediate disorder were the variances
still decrease with $L$, however the tails of the distributions acquire
more weight with a clear departure from a Gaussian form. Interestingly,
after rescaling the matrix elements by the standard deviation of their
distributions, we find that the distributions collapse reasonably
well for various $L$, not only for the Gaussian case, but also when
the distributions are non-Gaussian, indicating that the anomalous
behavior persists also in the thermodynamic limit \cite{SuppMat2018}.
In what follows we set $0.5\leq W\leq4$ and focus only on the ergodic,
albeit anomalous phase \cite{SuppMat2018}.

\begin{figure}
\includegraphics[width=1\columnwidth]{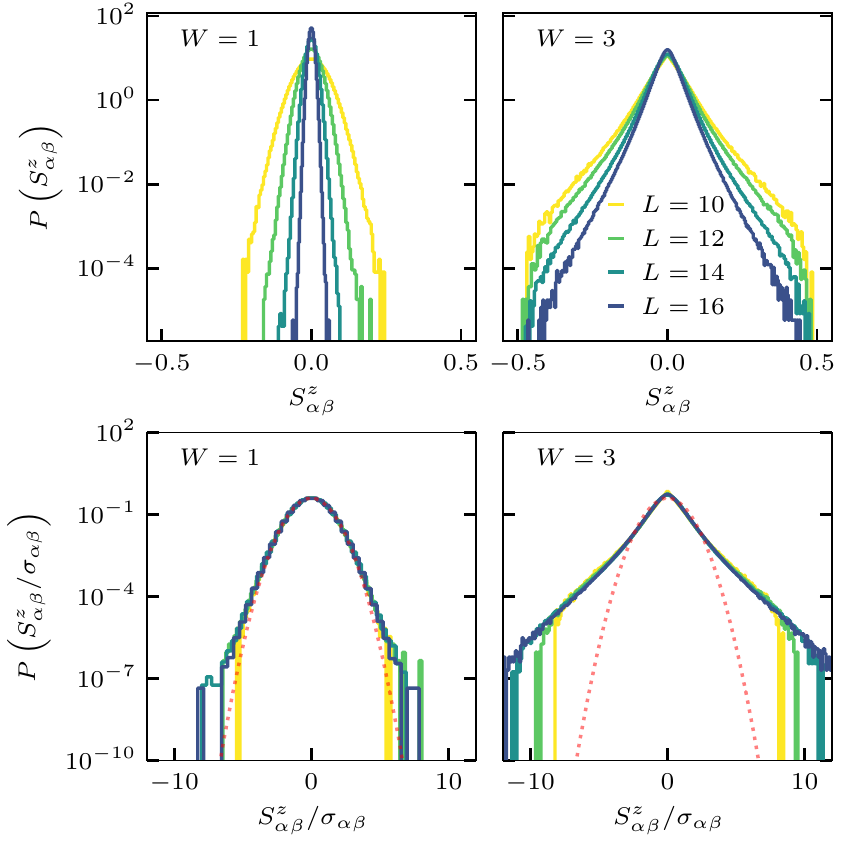}
\caption{\emph{Top row}: Probability distribution, $P\left(S_{\alpha\beta}^{z}\right)$
for $W=1$, $3$ and $L=10$, $12$, $14$ and $16$. \emph{Bottom}
\emph{row}: Same as top row, but with the distributions scaled with
the standard deviation of $P\left(S_{\alpha\beta}^{z}\right)$, $\sigma_{\alpha\beta}$.
The dashed red lines in the bottom row denote the standard normal
distribution. Other parameters are the same as in Fig.~\ref{fig:fig1-diag_matel_dist}.}
\label{fig:fig2-offdiag_matel_dist} 
\end{figure}
While the diagonal matrix elements of local operators are related
to the stationary state of the system, the offdiagonal matrix elements
are directly connected to thermalization, or the relaxation to the
stationary state. In Fig.~\ref{fig:fig2-offdiag_matel_dist} we show
the distributions of the offdiagonal matrix elements, $P\left(S_{\alpha\beta}^{z}\right)$
for two disorder values and various system sizes. To take statistics
over the entire quasienergy spectrum, for every Floquet eigenstate
we consider 10 Floquet eigenstates closest to it in quasienergy and
calculate $S_{\alpha\beta}^{z}$ for all the pairs in each group.The
matrix elements are then accumulated over all the groups as also over
different disorder realizations. For weak disorder $\left(W=1\right)$
the distributions are close to Gaussian in accordance with what is stipulted by ETH.
However for intermediate disorder, but still in the ergodic
phase $\left(W=3\right),$ the distributions are clearly non-Gaussian.
Rescaling the distributions by their standard deviation collapses
all system sizes on top of each other (see bottom panels of Fig.~\ref{fig:fig2-offdiag_matel_dist})
indicating the convergence of the shape of the distributions to their
thermodynamic limit. Similar anomalous behavior of the distributions
of the offidiagonal elements was observed by two of us for static
disordered systems, where it was established that only a \emph{modified}
version of the ETH is satisfied \cite{Luitz2016b}.

\begin{figure}[h]
\includegraphics[width=1\columnwidth]{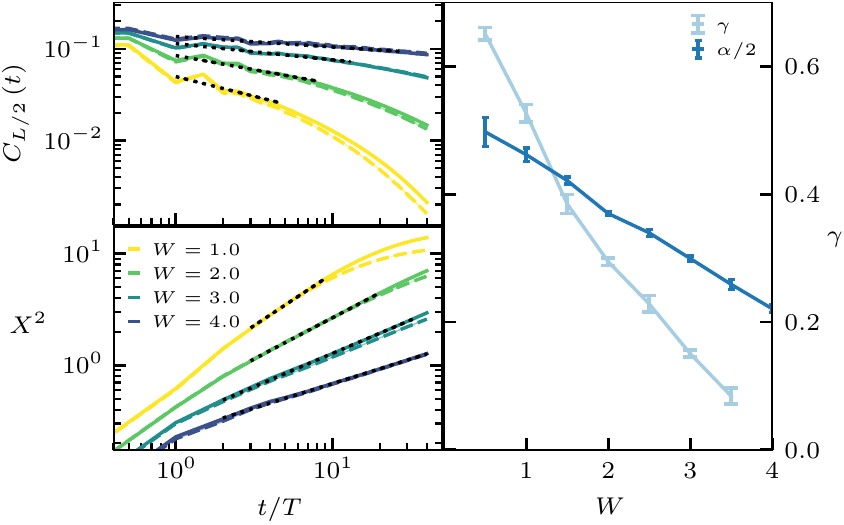}
\caption{\emph{Left}: Autocorrelation function, $C_{L/2}\left(t\right)$ (top)
and mean square displacement, $X^{2}\left(t\right)$ (bottom) as a
function of time on a log-log scale. Dashed lines, $L=23$ and solid
lines, $L=27$. Shades indicate statistical uncertainty (in most cases
smaller than the linewidth) and best fits of the underlying power
law are indicated by dotted black lines. \emph{Right}: The exponents
$\gamma$ and $\alpha/2$ as a function of the disorder strength $W$.
Error bars indicate only statistical errors and not systematic uncertainty.}
\label{fig:fig3-4-autocorr-msd} 
\end{figure}

\textit{Anomalous spin transport.\textemdash }Previous works on many-body
localization in static disordered systems identified a regime of anomalously
slow dynamics at the ergodic side of the MBL transition. In particular,
subdiffusive transport of spin or particles \cite{BarLev2014,Lev2014,Agarwal2014,BarLev2015,Znidaric2016}
as well as subballistic spreading of quantum information \cite{Luitz2015a,Luitz2017}
was observed. It was also shown that these dynamical properties are
related to a regime of slow anomalous thermalization described by
a \emph{modified} version of the ETH \cite{Luitz2016b}. For the driving
in Eq. (\ref{eq:hamfloquet}), the total magnetization is conserved,
which allows us to study spin transport. For this purpose we examine
the infinite temperature spin-spin correlation function 
\begin{equation}
C_{i}\left(t\right)=\frac{1}{\mathcal{N}}\tr\left(\hat{S}_{i}^{z}\left(t\right)\hat{S}_{L/2}^{z}\right),\label{eq:zzcorr}
\end{equation}
where $\mathcal{N}$ is the Hilbert space dimension and $\hat{S}_{i}^{z}\left(t\right)$
is the spin operator on site $i$ written in the Heisenberg picture
and evolved according to the aforementioned driving protocol. This
correlation function encodes the spreading of an initial magnetization
excitation created at center of the lattice, $L/2$, at time $t=0$.
To calculate (\ref{eq:zzcorr}) we exploit the concept of dynamical
typicality (for details see Sec 5.1.5 of Ref.~\cite{Luitz2016c}).
Practically, we approximate the trace in (\ref{eq:zzcorr}) by the
expectation value with respect to a random state, $\ket{\psi}$, sampled
according to the Haar measure. The error of this approximation is
inversely proportional to the square-root of the Hilbert space dimension.
After this substitution, the calculation of $C_{i}\left(t\right)=\bra{\psi}\hat{S}_{i}^{z}\left(t\right)\hat{S}_{L/2}^{z}\ket{\psi}$
can be reduced to the propagation of two wavefunctions according the
driving protocol in Eq.~\eqref{eq:hamfloquet}. The propagation is
performed using standard Krylov space time evolution methods (see
Sec. 5.1.2 in Ref. \cite{Luitz2016c}) for spin-1/2 chains of up to
27 spins. We characterize transport by the calculation of the spin-spin
autocorrelation function, $C_{L/2}\left(t\right)$ and the mean square
displacement (MSD),
\begin{equation}
X^{2}\left(t\right)=\sum_{i=1}^{L}\left(i-\frac{L}{2}\right){}^{2}\left[C_{i}\left(t\right)-C_{i}\left(0\right)\right],\label{eq:msd}
\end{equation}
which is directly related to the current-current correlation function
and therefore to transport (see Appendix of Ref.~\cite{Luitz2016c}).
The autocorrelation function of transported quantities decays as $C_{L/2}\left(t\right)\sim t^{-\gamma}$
and the MSD grows as, $X^{2}\left(t\right)\propto t^{\alpha}$. For
diffusive systems, $C_{i}\left(t\right)$ asymptotically assumes a
Gaussian form, yielding the connection, $\gamma=\alpha/2$, with $\alpha=1$.
For subdiffusive transport, $\alpha<1$, and typically $\gamma\neq\alpha/2$,
since $C_{i}\left(t\right)$ is non-Gaussian (cf. Fig.~\ref{fig:fig3-4-autocorr-msd}).
In the left panels of Fig.~\ref{fig:fig3-4-autocorr-msd} we present
the autocorrelation function and the MSD for various disorder strengths.
Both quantities are calculated by averaging the correlation function
$C_{i}\left(t\right)$ over 100-1000 disorder realizations. 
For weak disorder, we observe a fast decay of the autocorrelation
function and close to linear growth of the MSD, consistent with diffusive
transport observed in a similar clean Floquet model \cite{Luitz2017b}.
At intermediate disorder, we find a clearly sublinear growth of the
MSD with an exponent $\alpha<1$ which decreases as a function of
the disorder strength (see right panel of Fig. \ref{fig:fig3-4-autocorr-msd}).
The exponents were obtained by restricting the fits to times for which
finite size effects on the MSD are comparable to the statistical errors.

The analysis of the power-law decay of the autocorrelation function
is more involved, mostly due to the superimposed oscillations occurring
for short times. Fast transport (for smaller disorder strengths) results
in very short domain of power-law decay and less reliable exponents,
$\gamma$, for $W\leq1$. At stronger disorder, finite size effects
are less pronounced due to slower transport, yielding a longer domain
of the power laws and more reliable $\gamma$. We have verified that
the domain of the power law decay increases with increasing the system
size (see left top panel of Fig.~\ref{fig:fig3-4-autocorr-msd}). 

\begin{figure}
\includegraphics[width=1\columnwidth]{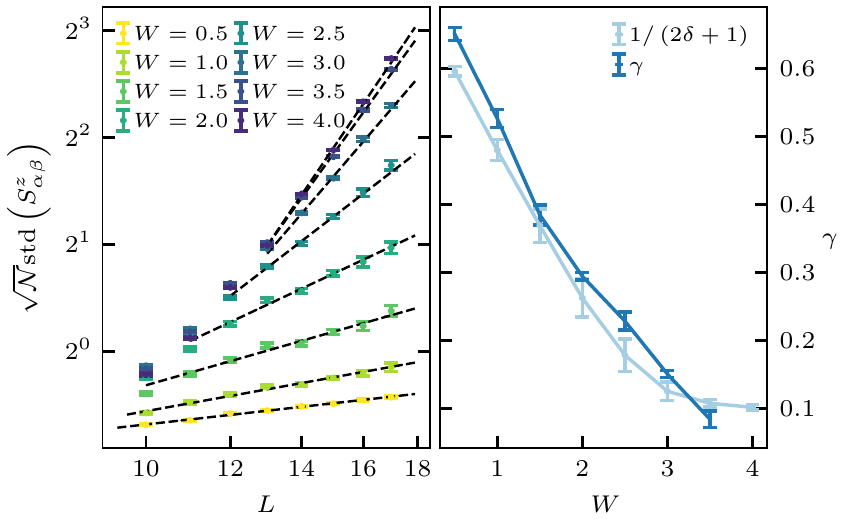} \caption{\emph{Left}: Extraction of the exponent $\delta$ from the scaling
of $\mathcal{\sqrt{N}}\text{std}\left(S_{\alpha\beta}^{z}\right)$
with $L$. \emph{Right}: Comparison of the exponent $\gamma$ to its
value extracted from the relation, $\gamma=1/\left(2\delta+1\right)$.}
\label{fig:fig5-exps} 
\end{figure}
\textit{Anomalous thermalization.\textemdash{}} A relation between
the scaling of the variance of the offdiagonal matrix elements and
the decay of the autocorrelation function was previously derived \cite{Luitz2016b}.
While ETH assumes $\mathcal{\sqrt{N}}\text{std}\left(S_{\alpha\beta}^{z}\right)=O\left(1\right)$,
for static systems it was shown that for energies, $\epsilon_{\alpha}$
and $\epsilon_{\beta}$ satisfying $\vert\epsilon_{\alpha}-\epsilon_{\beta}\vert<L^{-1/\gamma}$
the scaling is modified to $\mathcal{\sqrt{N}}\text{std}\left(S_{\alpha\beta}^{z}\right)\sim L^{\delta}$,
namely the decay of $\text{std}\left(S_{\alpha\beta}^{z}\right)$
with system size is not purely exponential, but includes a power-law
correction with system size. Moreover it was shown that $\delta=\left(1-\gamma\right)/\left(2\gamma\right)$
\cite{Luitz2016b}. While the derivation of this relation is rather
general, it was only tested for static systems. Here, we investigate
its validity for the Floquet system (\ref{eq:hamfloquet}). The exponent
$\gamma$ is obtained from the decay of the autocorrelation function
as explained in the previous section. The exponent $\delta$ is extracted
from the scaling of $\mathcal{\sqrt{N}}\text{std }\left(S_{\alpha\beta}^{z}\right)$
with system size (see left panel of Fig.~\ref{fig:fig5-exps}). To
calculate the standard deviation of $S_{\alpha\beta}^{z}$ we consider
approximately $10$ nearby (in quasienergy) Floquet eigenstates, $\ket{\alpha}$
and $\ket{\beta}$. Since the density of states is constant and exponentially
large in $L$, taking a fixed finite number of nearby states guarantees
that $\vert\epsilon_{\alpha}-\epsilon_{\beta}\vert<L^{-1/\gamma}$
is satisfied for sufficiently large systems. The extracted exponent
$\delta$ is nonzero, indicating a regime of anomalous (slow) thermalization
similar to the situation in static disordered systems \cite{Luitz2016b}.
The exponent $\delta$ is increasing with the disorder strength, presumably
diverging at the Floquet-MBL transition, were both ETH and its generalized
version fail. We note that the domain of validity of the power law
shifts to larger system sizes at intermediate disorder since the tails
of the distributions are only observable for large system sizes. The
right panel of Fig.~\ref{fig:fig5-exps} shows an excellent agreement
between the exponent $\gamma$, and its value calculated from $\delta$
using the relation $\delta=\left(1-\gamma\right)/\left(2\gamma\right)$. 

\textit{Discussion.\textemdash }In this Letter we have studied spin
transport and thermalization in an archetypal disordered and interacting
Floquet system, which has a Floquet-MBL transition for sufficiently
strong disorder. We have found, that similar to their static counterparts,
spin transport for disordered Floquet systems is subdiffusive and
is accompanied by anomalous thermalization with a modified form of
ETH. The distributions of matrix elements of local operators written
in the eigenbasis of the Floquet operator are non-Gaussian, although
the variance of these distributions still decay with the system size.
We demonstrated that the decay of the variance is directly related
to the temporal decay of the spin-spin autocorrelation function. Given
the above, we conjecture that the slow dynamical regime and anomalous
thermalization is a \emph{generic }feature of the ergodic phase of
systems exhibiting MBL, and does not rely on energy conservation.
It is interesting to see how removal of all conservation laws affects
on the dynamics of generic correlation functions in the system.

Interestingly, the disordered Floquest system we consider not only
has a flat many-body density of states, but also a structruless diagonal
elements of local operators written in the eigenbasis of the Floquet
operator. We argue that this finding is inconsistent with a mobility-edge
in the quasienergy spectrum, such that the Floquet-MBL transition
\emph{occurs at a critical disorder strength and has no additional
structure in quasienergy}. We leave a detailed comparison of the MBL
transition in static and Floquet systems for future studies. 
\begin{acknowledgments}
The authors gratefully acknowledge Achilleas Lazarides for useful
dicussions over the course of this project and also for related collaborations
on previous works on Floquet systems. This project has received funding
from the European Union's Horizon 2020 research and innovation programme
under the Marie Sk\l odowska-Curie grant agreement No. 747914 (QMBDyn)
and was in part supported by EPSRC Grant No. EP/N01930X/1. DJL acknowledges
PRACE for awarding access to HLRS's Hazel Hen computer based in Stuttgart,
Germany under grant number 2016153659. This work used the Extreme
Science and Engineering Discovery Environment (XSEDE), which is supported
by National Science Foundation Grant No. OCI-1053575.
\end{acknowledgments}

\bibliographystyle{apsrev4-1}
\bibliography{lib_yevgeny,local}

\begin{center}
\textbf{Supplementary material}
\end{center}

\emph{Level spacing ratio}.\textemdash In this section we provide
numerical evidence for the presence of a Floquet-ETH to Floquet-MBL
transition in the model considered via the commonly used diagnostic
of statistics of level spacing ratios of quasienergies.

As the quasienergies are defined only modulo $2\pi/T$, we choose
a particular branch namely $(-\pi/T,\pi/T]$ and within this branch
order the quasienergies such that $-\pi/T\le\epsilon_{1}<\epsilon_{2}<\cdots<\pi/T$.
The normalized level spacing ratio between consecutive levels is then
defined as 
\begin{equation}
r_{n}=\frac{\min\left(\delta_{n},\delta_{n-1}\right)}{\max\left(\delta_{n},\delta_{n-1}\right)}\qquad\delta_{n}=\epsilon_{n+1}-\epsilon_{n}.\label{eq:r}
\end{equation}

In the MBL phase, owing to the absence of level repulsion, the distribution
of $r$, $P(r)$ is expected to follow Poisson distribution $P(r)=2/(1+r)^{2}$. In the ergodic phase, $P(r)$ is expected
to be consistent with the circular unitary ensemble (CUE).

\begin{figure}
\includegraphics[width=1\columnwidth]{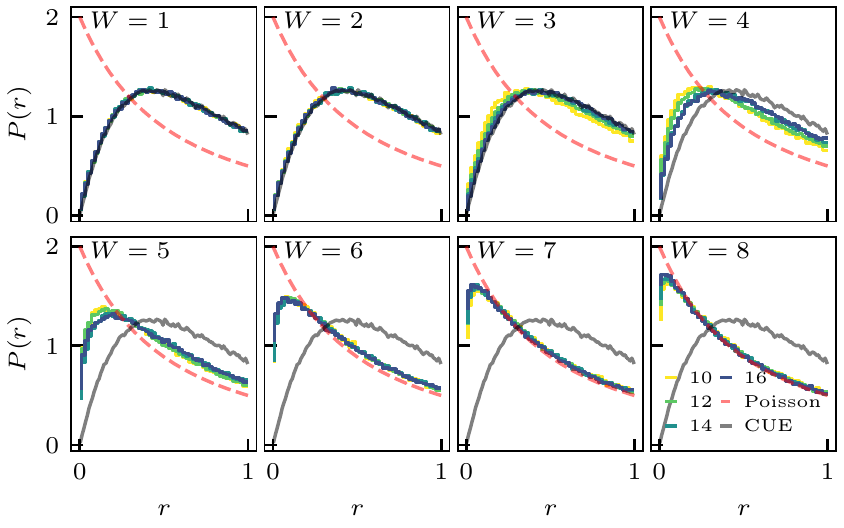}
\caption{The distributions of the level spacing ratios, $P(r)$, for various
disorder strengths, $W$, show an evolution from a CUE form at weak
disorder to a Poisson form at strong disorder. The theoretical curve
for CUE is obtained by numerically diagonalizing 5000 realizations
of random unitary matrices of dimension 200. }
\label{fig:fig8-LSRdists} 
\end{figure}

The numerical results for $P(r)$ are shown in Fig.~\ref{fig:fig8-LSRdists}
where we also show the theoretical curves corresponding to the cases
of CUE and Poisson statistics. We evaluate $P(r)$ for the CUE by
numerically diagonalizing large random unitary matrices. The results
show that at weak disorder, the distributions coincide with the CUE
result and at strong disorder with the Poisson result, confirming
the presence of the Floquet-MBL transition.

\begin{figure}
\includegraphics[width=1\columnwidth]{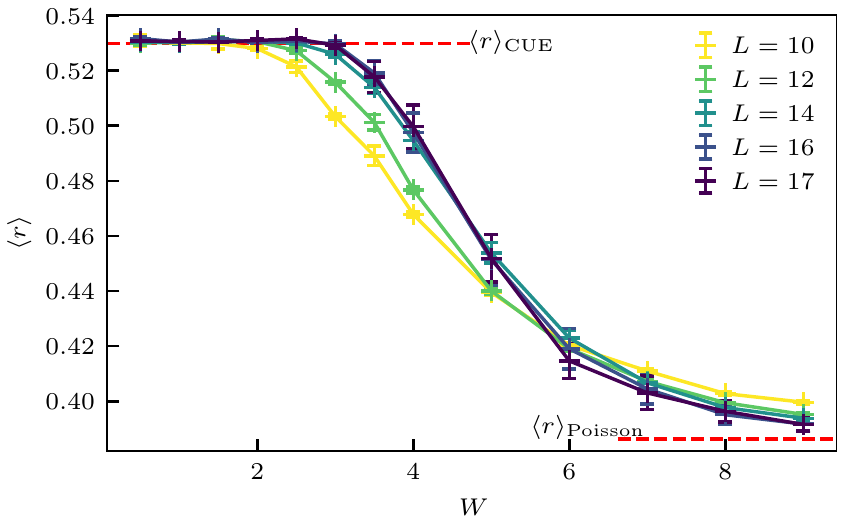}
\caption{The mean level spacing ratio, $\langle r\rangle$, shows a transition
from CUE to the Poisson value with increasing $W$, suggesting an
MBL transition. The apparent crossing for various system sizes, $L$,
albeit with a drift, suggests that $W<4$ is ergodic.}
\label{fig:fig7-meanLSR} 
\end{figure}

An approximate estimation of the critical disorder can be obtained
from the mean level spacing ratio, $\langle r\rangle$, which takes
the value of $\approx0.53$ for the CUE, and $2\log2-1\approx0.386$
for Poisson. While from the accessible system sizes, it is hard to
accurately estimate the critical disorder, the results shown in Fig.~\ref{fig:fig7-meanLSR}
suggest that for $W<4$, the phase is ergodic.

\begin{figure}
\includegraphics[width=1\columnwidth]{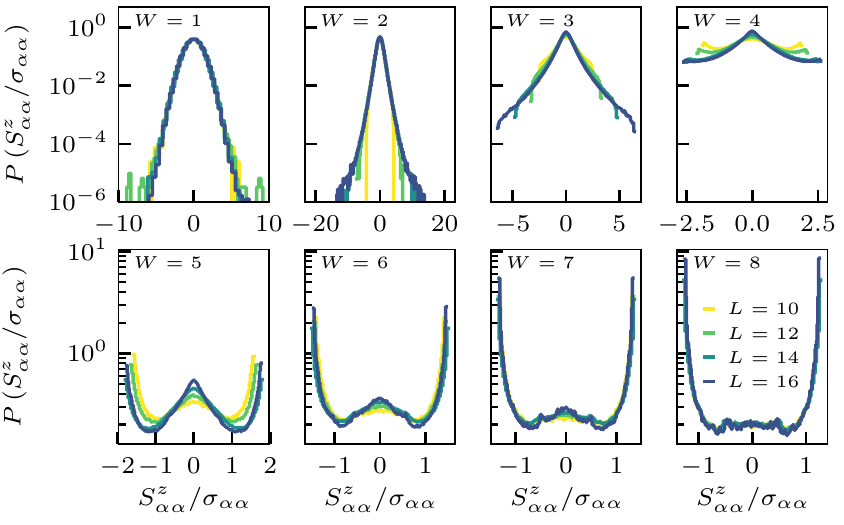}
\caption{The distributions of the scaled diagonal matrix elements, $P\left(S_{\alpha\alpha}^{z}/\sigma_{\alpha\alpha}\right)$,
collapse onto each other for various system sizes suggesting that
the form goes from a Gaussian at weak disorder to a bimodal distribution
at strong disorder, accompanied by an anomalous non-Gaussian regime
at intermediate disorder in the ergodic phase.}
\label{fig:fig6-diagdistscaled} 
\end{figure}

\emph{Distribution of diagonal matrix elements}.\textemdash In the
main text, the distribution of the diagonal matrix elements of a local
operator, $S_{\alpha\alpha}^{z}=\left\langle \alpha\left|\hat{S}_{\ell}^{z}\right|\alpha\right\rangle $,
written in the basis of Floquet eigenstates was used demonstrate the
behavior across the Floquet-MBL transition. The deviation of the distributions
$P\left(S_{\alpha\alpha}^{z}\right)$ from a Gaussian form at intermediate
disorder strengths, but still on the ergodic side, was indicative
of anomalous behavior. Here we provide additional evidence to the
existence of the ergodic and nonergodic phases and that the aforementioned
deviation from a Gaussian distribution is not just a finite size.
To show this we plot the distributions of $S_{\alpha\alpha}^{z}$
scaled with the standard deviation of $P\left(S_{\alpha\alpha}^{z}\right)$
denoted by $\sigma_{\alpha\alpha}$ for a number of $W$ as shown
in Fig.~\ref{fig:fig6-diagdistscaled}.

The evolution of the distributions from a Gaussian at weak disorder,
concomitant with ETH, to a bimodal distribution at strong disorder,
as expected for MBL, is clearly visible. At intermediate disorder
values, in the anomalous regime, $P\left(S_{\alpha\alpha}^{z}/\sigma_{\alpha\alpha}\right)$
for different system sizes seem to collapse onto each other, suggesting
a universal scaling form for the distribution which deviates from
a Gaussian even in the thermodynamic limit.

\begin{figure}[!h]
\includegraphics[width=1\columnwidth]{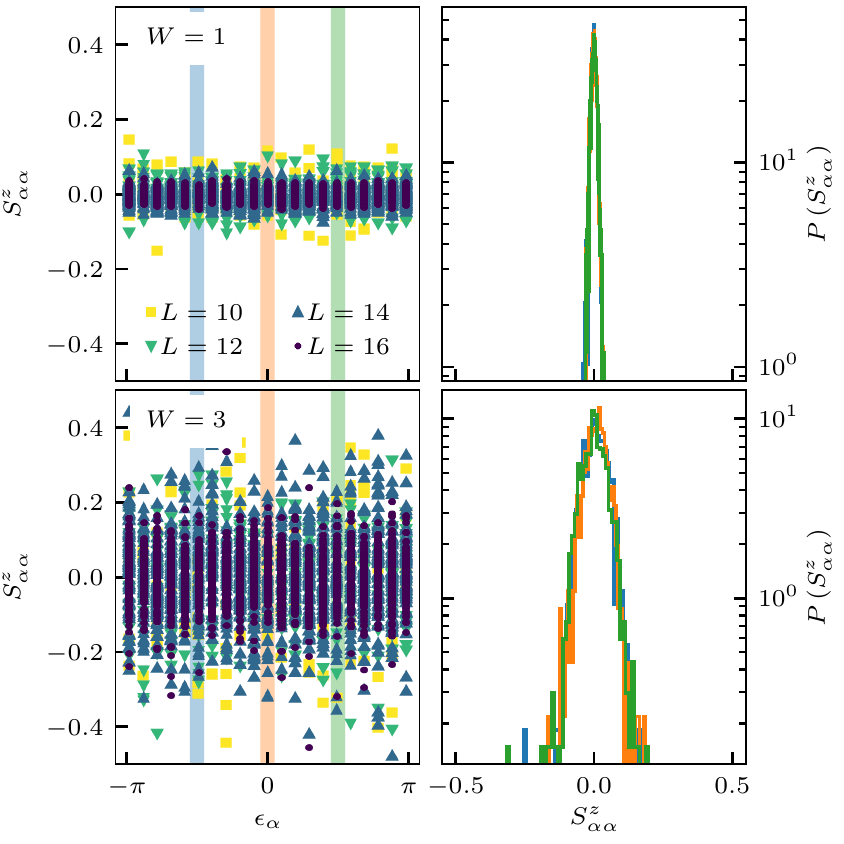}
\caption{Quasienergy resolved distributions $P\left(S_{\alpha\alpha}^{z}\right)$
for two values of disorder, $W=1$ and $W=3$. The left panels show
the eigenstate expectation values $S_{\alpha\alpha}^{z}$ at target
quasienergies $\epsilon=\epsilon_{\alpha}$ for various system sizes,
$L$, and just \emph{one} disorder realization. We ensure the equivalence
of the distributions over various system sizes by choosing the random
fields of the smaller system sizes, as a subset of the largest system
size, $16$,. We use the operator at site $\ell=2$ for all system
sizes. The right panels show the distributions for $L=16$ in three
different quasienergy windows shown by the vertical shaded region
in the left panels with the corresponding color.}
\label{fig:fig9-qen_resolved_eev} 
\end{figure}

In an undriven random-field XXZ chain, there exists a mobility edge
leading to a finite fraction of both, localized and delocalized eigenstates
at intermediate disorder. In the Floquet
system however, we find that this is not the case and at intermediate
disorder, all Floquet eigenstates are delocalized. We infer this from
studying the behavior of $S_{\alpha\alpha}^{z}$ as a function of
quasienergy $\epsilon$ as shown in Fig.~\ref{fig:fig9-qen_resolved_eev}.
The distribution of $S_{\alpha\alpha}^{z}$ is independent of the
quasienergy $\epsilon$, and for all $\epsilon$ shrinks with increasing
the system size, $L$, signifying that all states are delocalized.

\end{document}